\documentclass[12pt,eqno]{article}
\usepackage{amsmath,amssymb}
\numberwithin{equation}{section}

\newcommand{\alp}{\alpha}
\newcommand{\bt}{\beta}
\newcommand{\gm}{\gamma}

\newcommand{\dlt}{\delta}

\newcommand{\ep}{\epsilon}
\newcommand{\tht}{\theta}
\newcommand{\btht}{\bar{\tht}}

\newcommand{\kp}{\kappa}
\newcommand{\lmd}{\lambda}

\newcommand{\sgm}{\sigma}
\newcommand{\Sgm}{\Sigma}
\newcommand{\vph}{\varphi}

\newcommand{\Omg}{\Omega}

\newcommand{\ztR}{\zeta_{\rm R}}

\newcommand{\vth}{\vartheta^*}

\newcommand{\be}{\begin{equation}}
\newcommand{\ee}{\end{equation}}
\newcommand{\bea}{\begin{eqnarray}}
\newcommand{\eea}{\end{eqnarray}}
\newcommand{\eql}{\!\!\!&=\!\!\!&}

\newcommand{\defa}{\!\!\!&\equiv\!\!\!&}

\newcommand{\mtrx}[4]{\brkt{\begin{array}{cc}#1&#2\\#3&#4\end{array}}}
\newcommand{\dgnl}[2]{\brkt{\begin{array}{cc}#1& \\ &#2\end{array}}}
\newcommand{\vct}[2]{\brkt{\begin{array}{c}#1\\#2\end{array}}}

\newcommand{\tl}[1]{\tilde{#1}}
\newcommand{\bdm}[1]{{\mbox{\boldmath $#1$}}}

\newcommand{\diag}{{\rm diag}}
\newcommand{\der}{\partial}
\newcommand{\dr}{\!\!d}
\newcommand{\hc}{{\rm h.c.}}
\newcommand{\ie}{{\it i.e.}}

\newcommand{\vev}[1]{\langle #1 \rangle}

\newcommand{\brkt}[1]{\left( #1 \right)}
\newcommand{\brc}[1]{\left\{ #1 \right\}}
\newcommand{\sbk}[1]{\left[ #1 \right]}
\newcommand{\abs}[1]{\left| #1 \right|}

\renewcommand{\Re}{{\rm Re}}
\renewcommand{\Im}{{\rm Im}}

\newcommand{\cA}{{\cal A}}

\newcommand{\cF}{{\cal F}}
\newcommand{\cH}{{\cal H}}

\newcommand{\cL}{{\cal L}}
\newcommand{\cM}{{\cal M}}
\newcommand{\cN}{{\cal N}}
\newcommand{\cO}{{\cal O}}

\newcommand{\cV}{{\cal V}}
\newcommand{\cW}{{\cal W}}

\renewcommand{\ge}[2]{e_{#1}^{\;\;#2}}
\newcommand{\udl}[1]{\underline{#1}}
\newcommand{\nV}{n_{\rm V}}
\newcommand{\nH}{n_{\rm H}}
\newcommand{\dmx}{d_{\alp}^{\;\;\bt}}
\newcommand{\gey}{\vev{\ge{y}{4}}}
\newcommand{\SUu}{SU(2)_{\mbox{\scriptsize $\bdm{U}$}}}
\newcommand{\Usp}{USp(2,2\nH)}

\newcommand{\gc}{g_{\rm c}^0}
\newcommand{\gh}{g_{\rm h}^0}
\newcommand{\gv}{g_{\rm h}^1}
\newcommand{\GR}{G_{\rm R}}

\newcommand{\NP}[1]{{\it Nucl.~Phys.}~{\bf #1}}
\newcommand{\PL}[1]{{\it Phys.~Lett.}~{\bf #1}}

\newcommand{\PR}[1]{{\it Phys.~Rev.}~{\bf #1}}
\newcommand{\PRL}[1]{{\it Phys.~Rev.~Lett.}~{\bf #1}}
\newcommand{\PTP}[1]{{\it Prog.~Theor.~Phys.}~{\bf #1}}

\newcommand{\JH}[1]{{\it JHEP}~{\bf #1}}

\begin{document}

\begin{titlepage}
\null
\begin{flushright}
 {\tt hep-th/0511208}\\
KUNS-1996
\\
OU-HET 547/2005
\\
November, 2005
\end{flushright}

\vskip 2cm
\begin{center}
\baselineskip 0.8cm
{\LARGE \bf Consistent dimensional reduction of 
five-dimensional off-shell supergravity
}

\lineskip .75em
\vskip 2.5cm

\normalsize

{\large\bf Hiroyuki Abe}${}^1\!${\def\thefootnote{\fnsymbol{footnote}}
\footnote[1]{\it e-mail address:abe@gauge.scphys.kyoto-u.ac.jp}}
{\large\bf and Yutaka Sakamura}${}^2\!${\def\thefootnote{\fnsymbol{footnote}}
\footnote[2]{\it e-mail address:sakamura@het.phys.sci.osaka-u.ac.jp}}

\vskip 1.5em

${}^1${\it Department of Physics, Kyoto University, \\ 
Kyoto 606-8502, Japan}

\vskip 1.0em

${}^2${\it Department of Physics, Osaka University, \\ 
Toyonaka, Osaka 560-0043, Japan}

\vspace{18mm}

{\bf Abstract}\\[5mm]
{\parbox{13cm}{\hspace{5mm} \small
There are some points to notice in the dimensional reduction of 
the off-shell supergravity. 
We discuss a consistent way of the dimensional reduction 
of five-dimensional off-shell supergravity compactified on $S^1/Z_2$. 
There are two approaches to the four-dimensional effective action, 
which are complementary to each other. 
Their essential difference is the treatment of the compensator 
and the radion superfields. 
We explain these approaches in detail 
and examine their consistency. 
Comments on the related works are also provided. 
}}

\end{center}

\end{titlepage}

\clearpage

\section{Introduction}
Higher dimensional supergravity (SUGRA) appears in many researches of 
particle physics and cosmology, for example, as an effective theory 
of the superstring theory or as a setup of the braneworld scenario. 
For higher dimensional SUGRA on a compact space, some sort of 
dimensional reduction to lower dimensions such as the Kaluza-Klein reduction 
is necessary in order to discuss the low-energy physics. 
Consistent dimensional reduction of SUGRA in diverse dimensions 
is extensively discussed, e.g., in Refs.~\cite{Cvetic,Liu}. 

Among higher dimensional SUGRA, 
five dimensional supergravity (5D SUGRA) compactified on an orbifold~$S^1/Z_2$ 
has been thoroughly investigated since it is shown to appear 
as an effective theory of the strongly-coupled heterotic 
string theory~\cite{HW} compactified on a Calabi-Yau 3-fold. 
Especially, the Randall-Sundrum model~\cite{RS} is attractive 
as an alternative solution to the hierarchy problem, 
and its supersymmetric (SUSY) version is also discussed 
in many papers~\cite{GP,ABN}. 
The low-energy effective theories of such brane-world models 
are four dimensional (4D). 
In this paper, we will discuss the dimensional reduction of 5D SUGRA 
compactified on $S^1/Z_2$ to derive the 4D effective action, 
mainly focus on the matter couplings including the radion multiplet. 
In the following, we will concentrate ourselves on the case 
that the background saturates 
Bogomol'ny-Prasad-Sommerfield (BPS) bound~\cite{BPS}, \ie, 
preserves $\cN=1$ SUSY.\footnote{
$\cN=1$ SUSY denotes supersymmetry with four supercharges in this paper. } 

In the global SUSY case, the superfield formalism is useful 
because it makes SUSY manifest and we can systematically construct 
the action. 
Its counterpart in SUGRA is the superconformal gravity formulation, 
or off-shell SUGRA. (See, for example, Refs.~\cite{KU,KO,KO2,FKO}.)
Since the 4D effective theory has only $\cN=1$ SUSY, 
it is convenient to decompose each 5D superconformal multiplet 
into $\cN=1$ multiplets~\cite{KO}. 
In fact, we can rewrite the 5D off-shell SUGRA action 
in terms of $\cN=1$ multiplets 
in the language of 4D off-shell SUGRA~\cite{PST,AS}. 
Thus it seems possible to treat 5D off-shell SUGRA 
as if 4D off-shell SUGRA. 
In fact, the corresponding treatment is possible in the global SUSY case. 
A 5D supersymmetric action can be rewritten 
in terms of $\cN=1$ superfields on the $\cN=1$ superspace~\cite{AGW,MP},
and each 5D $\cN=1$ superfield can be expanded 
into infinite 4D superfields. 
Thus we can easily obtain a 4D superspace action 
by integrating the fifth dimension. 
Namely, we can perform the Kaluza-Klein (K.K.) dimensional reduction 
keeping the $\cN=1$ superfield structure. 
Unfortunately, the corresponding procedure cannot be performed 
in the SUGRA case. 
The main obstacle is the fact that 5D off-shell SUGRA has 
some multiplets that cannot be expanded into K.K. 4D multiplets. 
For example, it has the 5D compensator multiplet, 
which is not an independent degree of freedom. 
Its component fields are rewritten in terms of the physical matter fields 
after the superconformal gauge fixing. 
Therefore, we cannot perform the K.K. expansion for this multiplet 
keeping the $\cN=1$ structure manifest. 
We have to move to the on-shell action {\it before} 
the dimensional reduction in this approach. 
We will refer to this approach as the ``5D off-shell approach" 
in the following. 

From the 4D point of view, the radius of the orbifold 
is also a dynamical degree of freedom and behaves as a scalar field, 
which is called the {\it radion}. 
This belongs to a chiral multiplet for the preserved $\cN=1$ SUSY. 
Using this radion multiplet, we can construct the 4D off-shell action 
that is consistent with 5D SUGRA. 
This is the other way of deriving the 4D effective action. 
In contrast to the 5D off-shell approach, 
the resultant 4D action is expressed in the off-shell formalism. 
Thus we will refer to this as the ``4D off-shell approach" 
in the following. 

In this paper, we will explain these two approaches to the effective theory 
in detail and examine their consistency. 
Each approach has both advantage and drawback, 
and the two approaches are complementary to each other. 
We will also give a brief comment 
on the well-known works related to ours~\cite{MP,LS}. 

The paper is organized as follows. 
In the next section, we will explain the 5D off-shell approach 
after a brief review of the off-shell formalism of 5D SUGRA. 
Using a simple model, we will demonstrate the derivation of 
the effective action and show some problems of this approach. 
In Sect.~\ref{4Doffshell}, we will explain the 4D off-shell approach 
and derive the effective action expressed by superfields. 
The consistency with the 5D off-shell approach is also examined. 
Sect.~\ref{summary} is devoted to the summary. 
The definitions of the $\cN=1$~superfields we use in this paper 
are collected in Appendix~\ref{SF_GF}, and the explicit forms of 
the mode functions on the slice of AdS${}_5$ geometry are listed 
in Appendix~\ref{mf_HHc}.

\section{5D off-shell approach} \label{5Doffshell}
In this section, we will briefly review the off-shell formalism 
of 5D SUGRA and discuss the derivation of 
the 4D effective theory based on it. 
Throughout this paper, we will use $\mu,\nu,\cdots=0,1,2,3,4$ 
for the 5D world vector indices, and $m,n,\cdots =0,1,2,3$ 
for the 4D indices. 
The coordinate of the fifth dimension compactified 
on the orbifold~$S^1/Z_2$ is denoted as $y$ ($0\leq y \leq \pi R$). 
The corresponding local Lorentz indices are denoted 
by underbarred indices. 
Here we will focus on the case where 
the background preserves 4D Poincar\'{e} symmetry. 
Then the background metric can be written as~\footnote{The metric 
convention is chosen as $\eta_{\udl{m}\udl{n}}=\diag (1,-1,-1,-1)$. }
\be
 ds^2 = e^{2\sgm}\eta_{\udl{m}\udl{n}}dx^m dx^n-dy^2, 
\ee
where $\sgm(y)$ is a function of only $y$, which is determined by 
solving the Einstein equation. 

\subsection{5D Off-shell action}
The off-shell formulation of 5D SUGRA was provided 
by Refs.~\cite{KO,KO2,FKO}. 
This is very systematic and consistent way of constructing 
the 5D SUGRA action. 
For the purpose of deriving the 4D effective theory, 
it is convenient to decompose each 5D superconformal multiplet into 
$\cN=1$ multiplets~\cite{KO}. 
In fact we can rewrite the 5D SUGRA action in terms of $\cN=1$ multiplets 
in the language of 4D off-shell SUGRA~\cite{PST}, \ie, 
the $D$-term and $F$-term formulae of Ref.~\cite{KU}. 
Especially, in the case where we do not 
discuss the fluctuation modes of the gravitational multiplet, 
the action can be expressed on the $\cN=1$ superspace 
by using the following superfields~\cite{AS}. 

From the vector multiplet~$\cV^I$ ($I=0,1,\cdots,\nV$), we can construct 
$\cN=1$ vector and chiral superfields~$V^I$ and $\Phi_S^I$.\footnote{
For simplicity, we will consider only abelian gauge groups in this paper. }
From the hypermultiplets~$\cH^{\hat{\alp}}$ ($\hat{\alp}=0,1,\cdots,\nH$), 
we can construct a pair of chiral 
superfields~$(\Phi^{2\hat{\alp}+1},\Phi^{2\hat{\alp}+2})$. 
The orbifold $Z_2$ parity for each superfield is listed 
in Table.~\ref{Z2_parity}. 
\begin{table}[t]
\begin{center}
\begin{tabular}{|c||c|c|c|c||c|c|} \hline
\rule[-2mm]{0mm}{7mm} & $V^0$ & $\Phi_S^0$ & $V^{I\neq 0}$ & 
 $\Phi_S^{I\neq 0}$ & $\Phi^{2\hat{\alp}+1}$ & $\Phi^{2\hat{\alp}+2}$ 
 \\ \hline 
$Z_2$-parity & $-$ & $+$ & $+$ & $-$ & $-$ & $+$ \\ \hline
\end{tabular}
\end{center}
\caption{Orbifold parity for each superfield}
\label{Z2_parity}
\end{table}
The vector multiplet~$\cV^{I=0}$ corresponds to the graviphoton multiplet, 
and the hypermultiplet~$\cH^{\hat{\alp}=0}$ plays a role of 
the compensator multiplet.\footnote{
In this paper, we will consider the case of one compensator multiplet. 
An extension to the multi-compensator case is straightforward. }
In order to focus on the matter couplings, we will fix the gravitational 
fields to the following background values.\footnote{
we can always restore the dynamical gravitational modes by replacing 
$d^4\tht$- and $d^2\tht$-integrals in Eq.(\ref{5D_Sinv}) 
by the $D$- and $F$-term formulae of 4D superconformal gravity~\cite{KU}. 
}
\bea
 \vev{\ge{m}{\udl{n}}} \eql e^{\sgm}\dlt_m^{\;\;n}, \;\;\;\;\;
 \gey = \mbox{constant}, \nonumber\\
 \vev{\psi_\mu^i} \eql 0, 
\eea
where $\psi_\mu^i$ is the gravitino ($i=1,2$ is an $\SUu$-index.). 
We can always rescale the coordinate $y$ so that $\gey=1$, 
but we will keep the $\gey$-dependence in the following expressions 
for the readers to recall its origin. 

We can also introduce the 4D vector and chiral superfields, 
$U^A$ ($A=1,2,\cdots$) and $S^a$ ($a=1,2,\cdots$), 
on the boundaries of the orbifold. 

The explicit forms of these $\cN=1$ superfields 
in the notations of Refs.~\cite{KO,KO2} 
are collected in Appendix~\ref{SF_GF}. 

The 5D off-shell SUGRA action can be written as 
\bea
 S \eql \int\dr^5x\;\brkt{\cL_{\rm vector}+\cL_{\rm hyper}
 +\sum_{\vth=0,\pi}\cL_{\rm brane}^{(\vth)}\dlt(y-\vth R)}, \nonumber\\
 \cL_{\rm vector} \eql \sbk{\int\dr^2\tht\;
 \frac{3C_{IJK}}{2}\brc{i\Phi_S^I\cW^J\cW^K
 +\frac{1}{12}\bar{D}^2\brkt{V^ID^\alp\der_yV^J
 -D^\alp V^I\der_y V^J}\cW^K_\alp}+\hc} \nonumber\\
 &&-e^{2\sgm}\int\dr^4\tht\;V_E^{-2} C_{IJK}V_S^IV_S^JV_S^K, \nonumber\\ 
 \cL_{\rm hyper} \eql -2e^{2\sgm}\int\dr^4\tht\;
 V_E\dmx\bar{\Phi}^\bt\brkt{e^{-2igV^It_I}}^\alp_{\;\;\gm}\Phi^\gm 
 \nonumber\\
 && -e^{3\sgm}\sbk{\int\dr^2\tht\;
 \Phi^\alp\dmx\rho_{\bt\gm}\brkt{\der_y-2g\Phi_S^It_I}^\gm_{\;\;\dlt} 
 \Phi^\dlt+\hc},  \nonumber\\
 \cL_{\rm brane}^{(\vth)} \eql \brc{\int\dr^2\tht\;
 f^{(\vth)}_{AB}(S)\cW^A\cW^B+\hc} 
 -e^{2\sgm}\int\dr^4\tht\;
 \brkt{\abs{\Phi^{\alp=2}}^2}^{2/3}\exp\brc{-K^{(\vth)}(\bar{S},S,U)} 
 \nonumber\\
 &&+e^{3\sgm}\brc{\int\dr^2\tht\;
 \brkt{\Phi^{\alp=2}}^2 P^{(\vth)}(S)+\hc}, 
 \label{5D_Sinv}
\eea
where $C_{IJK}$ is a real constant tensor which is completely symmetric 
for the indices, 
$\rho_{\alp\bt}$ is an antisymmetric tensor defined as 
$\rho_{\alp\bt}\equiv i\sgm_2\otimes\bdm{1}_{\nH+1}$, 
and $\dmx$ is a metric of the hyperscalar space 
\be
 \dmx=\mtrx{\bdm{1}_2}{}{}{-\bdm{1}_{2\nH}}.  \label{def_dmx}
\ee
Note that the sign of the metric 
for the compensator superfields~($\Phi^{\alp=1},\Phi^{\alp=2}$) is opposite 
to that for the matter superfields~$\Phi^{\alp}$ ($\alp\geq 3$). 
The summations for the indices are implicit. 
The generators~$t_I$ ($I=0,1,\cdots,\nV$) are defined as 
anti-hermitian. 
The superfield~$V_E$ corresponds to the $\cN=1$ real general 
multiplet~$\bdm{W}_y$ in Eq.(4.12) of Ref.~\cite{KO}, 
and is a spurion-like superfield 
since the gravitational fields are frozen here. 
\be
 V_E = \vev{\ge{y}{4}}-\tht^2\cF_V-\btht^2\bar{\cF}_V, 
\ee
where $\cF_V$ is an auxiliary field. 
The superfields~$V_S^I$ and $\cW_\alp^I$ are gauge-invariant quantities 
defined as 
\bea
 V_S^I \defa -\der_y V^I-i\Phi_S^I+i\bar{\Phi}_S^I, \nonumber\\
 \cW_\alp^I \defa -\frac{1}{4}\bar{D}^2D_\alp V^I. 
\eea
The first line of $\cL_{\rm vector}$ corresponds to the gauge kinetic terms 
and the supersymmetric Chern-Simons terms. 
In fact, the variation of these terms under the gauge transformation 
does not vanish, but becomes total derivative. 
Thus these terms cannot be expressed by only the above gauge-invariant quantities. 

In the boundary action, $f^{(\vth)}_{AB}$, $K^{(\vth)}(\bar{S},S,U)$ and 
$P^{(\vth)}(S)$ are the gauge kinetic function, the K\"{a}hler potential 
and the superpotential, respectively. 
$S^a$ and $U^A$ can be either superfields localized on the boundaries 
or induced from the bulk superfields. 
$\cW^A$ is a superfield strength of $U^A$. 
Note that only $\Phi^{\alp=2}$ can appear in the boundary actions 
as a chiral compensator superfield because $\Phi^{\alp=1}$ is odd 
under the $Z_2$-parity and vanishes on the boundaries. 
The powers of $\Phi^{\alp=2}$ are determined by the Weyl weight counting. 
Note also that the Weyl weights of the matter multiplets 
in the 4D off-shell action must be zero. 
Thus, if $S$ and $U$ are the induced superfields 
from the $Z_2$-even 5D superfields~$\Phi^{2\hat{\alp}+2}$ 
($\hat{\alp}\geq 1$) and $V^I$ ($I\neq 0$), they are identified as 
\bea
 S^{a=\hat{\alp}} \eql \frac{\Phi^{2\hat{\alp}+2}}{\Phi^{\bt=2}}, 
 \;\;\; (\hat{\alp}\geq 1) \nonumber\\
 U^{A=I} \eql V^I. \;\;\; (I \neq 0) \label{induced_S}
\eea

Eq.(\ref{5D_Sinv}) reproduces the off-shell SUGRA action in Ref.~\cite{KO2}. 
(See Appendix.~B of Ref.~\cite{AS}.)
Note that Eq.(\ref{5D_Sinv}) is just a shorthand expression 
for the full SUGRA action. 
We can always restore the gravitational fields to the dynamical ones 
by promoting $d^4\tht$- and $d^2\tht$-integrals 
to the $D$- and $F$-term formulae of 4D superconformal gravity~\cite{KU} 
and regarding each superfield as the corresponding $\cN=1$ multiplet.\footnote{
The explicit forms of $D$- and $F$-term formulae are compactly listed in 
Appendix~C of Ref.~\cite{KO}. }  
(See also Ref.~\cite{PST}.) 
Thus, we will use both words ``$\cN=1$ superfield" and ``$\cN=1$ multiplet" 
without clear distinction in the following.

\subsection{Superconformal gauge fixing and practical form of 5D action}
In order to obtain the Poincar\'{e} supergravity, we have to fix 
the extraneous superconformal symmetries. 
The required gauge-fixing conditions are listed in Appendix~\ref{scGF}. 
Here we will rewrite those conditions in our notation. 

We will denote each component of the superfields as~\footnote{
We have defined $\tht$- and $\tht^2$ components of 
the chiral superfields with minus signs 
in order to match the notation of Refs.~\cite{KO,KO2}. }
\bea
 V^I \eql \tht\sgm^{\udl{m}}\btht W_m^I+i\tht^2\btht\bar{\lmd}^I
 -i\btht^2\tht\lmd^I+\frac{1}{2}\tht^2\btht^2 D^I, \nonumber\\
 \Phi_S^I \eql \vph_S^I-\tht\chi_S^I-\tht^2\cF_S^I, \nonumber\\
 \Phi^\alp \eql \vph^\alp-\tht\chi^\alp-\tht^2\cF^\alp, 
\eea
where $I=0,1,\cdots,\nV$ and $\alp=1,2,\cdots,2(\nH+1)$. 

In terms of these component fields, the gauge-fixing conditions can be 
rewritten as follows. 
\bea
 \vph^{\alp=1} \eql 0, \nonumber\\
 \vph^{\alp=2} \eql \brkt{M_5^3+\sum_{\bt=3}^{2\nH+2}\abs{\vph^\bt}^2}^{1/2} 
 = M_5^{3/2}+\frac{\kp^{3/2}}{2}
 \sum_{\bt=3}^{2\nH+2}\abs{\vph^\bt}^2+\cO(\kp^{9/2}), 
 \nonumber\\
 \chi^{\alp=1} \eql -\brkt{\vph^{\alp=2}}^{-1}\sum_{\bt,\gm=3}^{2\nH+2}
 \rho_{\bt\gm}\vph^\bt\chi^\gm, 
 \nonumber\\
 \chi^{\alp=2} \eql \brkt{\vph^{\alp=2}}^{-1}\sum_{\bt=3}^{2\nH+2}
 \bar{\vph}^\bt\chi^\bt, \nonumber\\ 
 \Im\vph_S^{I=0} \eql \frac{\gey M_5}{2}
 +\frac{\kp}{3\gey}\sum_{J=1}^{\nV}\brkt{\Im\vph_S^J}^2+\cO(\kp^5), 
 \nonumber\\
 \lmd^{I=0} \eql \frac{2\kp}{3\gey}\sum_{J=1}^{\nV}
 \brkt{\Im\vph_S^J}\lmd^J+\cO(\kp^{9/2}), \nonumber\\
 \chi_S^{I=0} \eql \frac{2\kp}{3\gey}\sum_{J=1}^{\nV}
 \brkt{\Im\vph_S^J}\chi_S^J+\cO(\kp^{9/2}), 
 \label{GF_cond}
\eea
where $M_5$ is the 5D Planck mass and $\kp\equiv 1/M_5$. 
For simplicity, we have assumed that 
the vector sector is maximally symmetric.\footnote{
Extension to more generic case, for example the cases discussed 
in Refs.~\cite{Cvetic,Liu}, will be straightforwardly possible. 
}
Namely,  
\be
 C_{IJK}M^IM^JM^K = (M^{I=0})^3-\frac{1}{2}M^{I=0}
 \sum_{J=1}^{\nV}(M^J)^2,  \label{sym_cN}
\ee
where $M^I$ is the gauge scalar in the 5D vector multiplet~$\cV^I$. 
(See Appendix~\ref{5Dsf}.) 
If we want to include the Fayet-Iliopoulos term, 
the above cubic function of $M^I$ is corrected 
by adding an extra term~\cite{AK}. 

Therefore, the compensator multiplet~$\Phi^{\alp=1,2}$ and 
the graviphoton multiplet~($V^0,\Phi_S^0$) are not independent fields 
except for their auxiliary fields and the graviphoton field~$W_\mu^0$. 
In order to discriminate these multiplets from the other physical 
multiplets, we will rewrite the superfields as follows.\footnote{
Here we consider the case of $\nH=\nV=1$ case. 
The extension to the case of arbitrary $\nH$ and $\nV$ 
is straightforward. }
For the hypermultiplets, 
\bea
 \Xi \defa \kp^{3/2}\Phi^{\alp=2} = \xi-\tht\chi_\xi-\tht^2\cF_\xi, \nonumber\\
 \Xi^c \defa \kp^{3/2}\Phi^{\alp=1} = \xi^c-\tht\chi_\xi^c
 -\tht^2\cF_\xi^c, \nonumber\\
 H \defa \sqrt{2}\Phi^{\alp=4} = h-\tht\chi_h-\tht^2\cF_h, \nonumber\\
 H^c \defa \sqrt{2}\Phi^{\alp=3} = h^c-\tht\chi_h^c-\tht^2\cF_h^c. 
\eea
For the vector multiplets, 
\bea
 V^0 \defa V^{I=0} =\tht\sgm^{\udl{m}}\btht W^0_m+i\tht^2\btht\bar{\lmd}^0
 -i\btht^2\tht\lmd^0+\frac{1}{2}\tht^2\btht^2 D^0, \nonumber\\
 E \defa -2i\kp\Phi_S^{I=0} = \vph_E-\tht\chi_E-\tht^2\cF_E, \nonumber\\
 V \defa \sqrt{\frac{M_5}{2}}V^{I=1} = \tht\sgm^{\udl{m}}\btht W_m
 +i\tht^2\btht\bar{\lmd}-i\btht^2\tht\lmd+\frac{1}{2}\tht^2\btht^2D,  
 \nonumber\\
 \Phi_S \defa \sqrt{\frac{M_5}{2}}\Phi_S^{I=1} = \vph_S
 -\tht\chi_S-\tht^2\cF_S. 
\eea
Then, Eq.(\ref{GF_cond}) is rewritten as 
\bea
 \xi \eql \brc{1+\frac{\kp^3}{2}\brkt{\abs{h}^2+\abs{h^c}^2}}^{1/2}, 
 \;\;\;\;\; 
 \xi^c = 0, \nonumber\\
 \chi_\xi \eql \frac{\kp^{3/2}}{2\xi}
 \brkt{\bar{h}\chi_h+\bar{h}^c\chi_h^c}, 
 \;\;\;\;\;
 \chi_\xi^c = -\frac{\kp^{3/2}}{2\xi}\brkt{h\chi_h^c-h^c\chi_h}, 
 \nonumber\\
 \lmd^0 \eql \frac{4\kp^2}{3\gey}\brkt{\Im\vph_S}\lmd+\cO(\kp^{9/2}), \nonumber\\
 \vph_E \eql \gey-i\kp W_y^0, 
 \;\;\;\;\;
 \chi_E = -\frac{8i\kp^3}{3\gey}\brkt{\Im\vph_S}\chi_S+\cO(\kp^{9/2}). 
\eea
We have considered the maximally symmetric case in the vector sector, 
\ie, $C_{IJK}$ is assumed as Eq.(\ref{sym_cN}). 

The compensator and the physical hypermultiplets can be 
charged for the graviphoton~$W_\mu^0$ and 
the ordinary gauge field~$W_\mu$. 
Here we will choose the directions of these gaugings to 
the $\sgm_3$-direction because the gauging along the other direction 
mixes $\Phi^{2\hat{\alp}+1}$ and $\Phi^{2\hat{\alp}+2}$, 
which have opposite $Z_2$-parities. 
Thus, the anti-hermitian generator~$t_0$ and $t_1$ are 
\be
 gt_0 = -i\dgnl{\gc}{\gh}\otimes \sgm_3, \;\;\;\;\;
 gt_1 = -i\dgnl{0}{\gv}\otimes \sgm_3.  
\ee
The Pauli matrix~$\sgm_3$ acts on each hypermultiplet 
$(\Phi^{2\hat{\alp}+1},\Phi^{2\hat{\alp}+2})$. 
We have assumed that the compensator multiplet is neutral for $W_\mu$, 
for simplicity. 
Note that the couplings~$\gc$ and $\gh$ are $Z_2$-odd under 
the orbifold parity since the graviphoton~$W_m^0$ is odd. 
Such $Z_2$-odd couplings can be consistently introduced 
in the off-shell SUGRA action by the method proposed 
in Refs.~\cite{FKO,BKV}. 

Then, the Lagrangians in Eq.(\ref{5D_Sinv}) are rewritten as 
\bea
 \cL_{\rm vector} \eql 
 \sbk{\int\dr^2\tht\;\brc{-\frac{3M_5}{4}E\brkt{\cW^0}^2
 +\frac{1}{4}E\cW^2-\frac{i}{2}\Phi_S \cW^0\cW+C(V^0,V)}+\hc} 
  \nonumber\\
 && -e^{2\sgm}\int\dr^4\tht\; 
 V_E^{-2}\left\{\brkt{-\der_y V^0+\frac{M_5}{2}(E+\bar{E})}^3 \right.
 \nonumber\\
 &&\hspace{30mm}\left.
 -\kp\brkt{-\der_y V^0+\frac{M_5}{2}(E+\bar{E})}
 \brkt{-\der_y V-i\Phi_S+i\bar{\Phi}_S}^2\right\} \nonumber\\
 \cL_{\rm hyper} \eql e^{2\sgm}\int\dr^4\tht\;V_E\left\{
 -2M_5^3\brkt{\bar{\Xi}e^{3\kp k\ep V^0}\Xi
 +\bar{\Xi}^ce^{-3\kp k\ep V^0}\Xi^c} \right. \nonumber\\
 &&\hspace{28mm}\left. 
 +\bar{H}e^{2\kp m\ep V^0+2gV}H
 +\bar{H}^ce^{-2\kp m\ep V^0-2gV}H^c\right\} \nonumber\\
 &&+e^{3\sgm}\left[\int\dr^2\tht\;\left\{-2M_5^3\Xi^c
 \brkt{\der_y+\frac{3}{2}\dot{\sgm}+\frac{3}{2}k\ep E}\Xi 
 \right.\right.\nonumber\\
 &&\hspace{25mm}\left.\left. 
 +H^c\brkt{\der_y+\frac{3}{2}\dot{\sgm}+m\ep E-2ig\Phi_S}H\right\}
 +\hc\right],  \nonumber\\
 \cL_{\rm brane}^{(\vth)} \eql 
 \brc{\int\dr^2\tht\;f^{(\vth)}_{AB}(S)\cW^A\cW^B+\hc} 
 -e^{2\sgm}M_5^2\int\dr^4\tht\;
 \brkt{\abs{\Xi}^2}^{2/3}\exp\brc{-K^{(\vth)}(\bar{S},S,U)} 
 \nonumber\\
 &&+e^{3\sgm}\brc{\int\dr^2\tht\;\Xi^2 P^{(\vth)}(S)+\hc},  
 \label{MO_action1}
\eea
where $\dot{\sgm}\equiv d\sgm/dy$, $k\ep\equiv 2\gc M_5/3$, 
$m\ep\equiv \gh M_5$ and $g\equiv\sqrt{2\kp}\gv$.  
Here, $\ep$ is a periodic step function defined as 
\be
 \ep(y) = \left\{\begin{array}{l} +1 \;\;\;\; (2n\pi R < y < (2n+1)\pi R) \\
 -1 \;\;\;\; ((2n-1)\pi R < y < 2n\pi R). \;\;\;\; (\mbox{$n$: integer}) 
 \end{array}\right.  \label{def_ep}
\ee
Note that the coupling constants~$\gc$ and $\gh$ are $Z_2$-odd. 
$C(V^0,V)$ corresponds to the second terms in the first line of 
$\cL_{\rm vector}$ in Eq.(\ref{5D_Sinv}), which is irrelevant to 
the following discussion. 
We have rescaled $P^{(\vth)}$ in Eq.(\ref{MO_action1}) as 
$P^{(\vth)}\to \kp^3 P^{(\vth)}$ from that in Eq.(\ref{5D_Sinv}).

\subsection{Simple model} \label{SUSY_stblz}
To see the role of the superfields $(\Xi,\Xi^c,V^0,E,V_E)$, 
we will consider a simple model proposed in Ref.~\cite{MO}. 
This corresponds to a supersymmetric extension of the Goldberger-Wise 
model~\cite{GW} that stabilizes the size of the fifth dimension 
by the bulk scalar field. 

The model consists of only one physical hypermultiplet~$(H, H^c)$ 
with the bulk mass~$m$ besides the compensator and the graviphoton 
multiplets. 
Since $H$ is $Z_2$-even, it can appear in the boundary actions. 
In this model, the following tadpole superpotentials are introduced 
on the boundaries. 
\be
 P^{(0)}\brkt{\frac{H}{\Xi}} = J_0\frac{H}{\Xi}, \;\;\;\;\;
 P^{(\pi)}\brkt{\frac{H}{\Xi}} = -J_\pi\frac{H}{\Xi}, 
\ee 
where $J_0,J_\pi$ are real positive constants. 
Recall that $H$ must appear in the boundary action in the form of 
$H/\Xi$. 
(See Eq.(\ref{induced_S}).)

Due to these boundary superpotentials, the $Z_2$-odd field~$h^c$ 
must satisfy the following boundary conditions. 
\be
 \sbk{h^c}_0 = J_0, \;\;\;\;\;
 \sbk{h^c}_\pi = J_\pi,  \label{hc_bd_cond}
\ee
where the symbol~$\sbk{\cdots}$ is defined as 
\be
 \sbk{\vph}_{\vth}\equiv \vph(y=\vth R+0)-\vph(y=\vth R-0). 
\ee

In the following, we will neglect the 4D component of the graviphoton 
$W_m^0$.\footnote{
Since $W_m^0$ is $Z_2$-odd, it does not have a zero-mode. }
Then, $V^0$ has only the auxiliary field~$D^0$ 
because $\lmd^0=0$ in our case, \ie, $(\nV,\nH)=(0,1)$. 

From the equation of motions for the auxiliary fields, 
we obtain~\footnote{They are obtained by taking the variation of 
$D^0$, $\cF_E$, $\cF_V$, $\cF_\xi$, $\cF_\xi^c$, $\cF_h$ and $\cF_h^c$, 
respectively. } 
\bea
 D^0 \eql -e^{2\sgm}M_5\brc{\dot{\sgm}+k\ep\abs{\xi}^2
 -\frac{\kp^3m\ep}{3}\brkt{\abs{h}^2-\abs{h^c}^2}}, \nonumber\\
 \cF_E \eql 2\cF_V-\frac{2}{3}e^\sgm\kp^3m\ep\bar{h}^c\bar{h}, 
 \nonumber\\
 \cF_E \eql 2\cF_V+\frac{2}{3}\cF_\xi\bar{\xi}
 -\frac{\kp^3}{3}\brkt{\cF_h\bar{h}+\cF_h^c\bar{h}^c}, \nonumber\\
 \cF_\xi \eql -\cF_V\xi-\frac{\kp^3}{2}\sum_{\vth=0,\pi}
 e^\sgm e^{i\vth}J_{\vth}\bar{h}\dlt(y-\vth R), \nonumber\\
 \cF_\xi^c \eql e^\sgm\brkt{\der_y+\frac{3}{2}\dot{\sgm}
 +\frac{3}{2}k\ep}\bar{\xi}, 
 \nonumber\\
 \cF_h \eql -\cF_V h-e^\sgm\brkt{
 \der_y+\frac{3}{2}\dot{\sgm}-m\ep}\bar{h}^c
 +\sum_{\vth}e^\sgm\bar{\xi}e^{i\vth}J_{\vth}\dlt(y-\vth R), 
 \nonumber\\
 \cF_h^c \eql -\cF_V h^c+e^\sgm\brkt{
 \der_y+\frac{3}{2}\dot{\sgm}+m\ep}\bar{h}. 
 \label{aux_EOM}
\eea

We will search the background that preserves $\cN=1$ SUSY. 
The BPS condition can be expressed as~\footnote{
We can show that this is equivalent to the Killing spinor equations, 
which is the conditions that the SUSY variations of all fermionic fields 
are zero. } 
\be
 \vev{D^0}=\vev{\cF_E}=\vev{\cF_V}=\vev{\cF_\xi}
 =\vev{\cF_\xi^c}=\vev{\cF_h}=\vev{\cF_h^c}=0. 
\ee

Solving this condition, we obtain the following BPS background. 
\bea
 \sgm(y) \eql -k\abs{y}-\frac{l^2}{24}e^{(3k+2m)\abs{y-\pi R}}+\cO(l^4), \nonumber\\
 h_{\rm cl} \eql 0, \nonumber\\
 h_{\rm cl}^c \eql \frac{J_0}{2}\ep(y)e^{\brkt{\frac{3}{2}k+m}\abs{y}}
 \brc{1+\cO(l^2)}, 
 \label{bkgd}
\eea
where $l\equiv\kp^{3/2}J_\pi$ is a dimensionless constant, 
which parametrizes the backreaction on the metric 
from the nontrivial configuration of the bulk scalar~$h^c$. 
We will assume that $J_0,J_\pi \ll M_5^{3/2}$, \ie, $l \ll 1$ in the following. 
We have used the first condition of Eq.(\ref{hc_bd_cond}) 
to determine the normalization of $h_{\rm cl}^c$. 
From the second condition of Eq.(\ref{hc_bd_cond}), we obtain 
\be
 J_0 = J_\pi e^{-\brkt{\frac{3}{2}k+m}\pi R}. 
\ee
Therefore, the radius~$R$ is stabilized with a finite value. 

The physical fields~$(h,\chi_h)$, $(h^c,\chi_h^c)$ are expanded into 
K.K. modes around the background~(\ref{bkgd}). 
First, let us consider the limit of $l\to 0$. 
In this case, $h_{\rm cl}^c$ vanishes, 
and the background geometry becomes a slice of AdS${}_5$. 
The radius~$R$ is not stabilized and the radion is massless. 
The radion mode resides 
only in the fluctuation of the f\"{u}nfbein~$\ge{\mu}{\udl{\nu}}$,
which we have dropped here. 
Since $H$ and $H^c$ are even and odd under the $Z_2$-parity, 
only $(h,\chi_h)$ have zero modes. 

Now we will turn on the parameter~$l$. 
Then $h^c$ has a nontrivial background configuration and stabilizes the radius. 
In this case, the radion obtains a nonzero mass of order $l$, 
and has its component also in the fluctuation of $h^c$ 
although the most part is still in the f\"{u}nfbein. 
The zero modes of $(h,\chi_h)$ also obtain nonzero masses of 
$\cO(l)$.\footnote{
The $Z_2$-odd fields~$(h^c,\chi_h^c)$ also have light modes 
whose masses are $\cO(l)$ in this case. 
In fact, they correspond to the radion mode and its superpartner. 
As explained in Sect.~2.2 of Ref.~\cite{AS2}, 
their normalized mode-functions are suppressed by $l$, 
and thus  we will neglect these modes in the following. }
We will calculate these masses in the following. 
The hyperscalars are mode-expanded as 
\bea
 h(x,y) \eql C_h e^{\brkt{\frac{3}{2}k-m}\abs{y}}\brc{1+\cO(l^2)}h_{(0)}(x)
 +(\mbox{massive modes}), \nonumber\\
 h^c(x,y) \eql h_{\rm cl}^c(y)+(\mbox{massive modes}), 
 \label{hhc_exd}
\eea
where the normalization factor~$C_h$ is defined as Eq.(\ref{def_Ch}). 

Substituting this background into Eq.(\ref{aux_EOM}), 
the on-shell values of the auxiliary fields are
\bea
 \cF_E \eql \frac{2}{3}\kp^3mJ_0\ep^2(y)C_he^{2k\abs{y}}\bar{h}_{(0)} 
 -\kp^3J_0C_h\sum_{\vth=0,\pi}e^{i\vth}e^{2k\abs{y}}\bar{h}_{(0)} 
 \dlt(y-\vth R)+\cdots, \nonumber\\
 \cF_V \eql \frac{\kp^3}{2}mJ_0\ep^2(y)C_he^{2k\abs{y}}\bar{h}_{(0)}
 -\frac{\kp^3}{2}J_0C_h\sum_{\vth=0,\pi}e^{i\vth}e^{2k\abs{y}}
 \bar{h}_{(0)}\dlt(y-\vth R)+\cdots, \nonumber\\
 \cF_\xi \eql -\frac{\kp^3}{2}mJ_0\ep^2(y)C_he^{2k\abs{y}}\bar{h}_{(0)}
 +\cdots, 
 \label{on-shell}
\eea
where the ellipses denote terms involving massive modes 
or higher order terms for $l$-expansion. 
The on-shell values of the other auxiliary fields start 
from the quadratic terms for $h_{(0)}$. 
Note that $\ep^2(y)$ equals one in the bulk, but its boundary values 
depend on the regularization of the orbifold singularity. 

Eliminating the auxiliary fields by Eq.(\ref{on-shell}) 
and performing the $y$-integral, 
we can calculate the mass of $h_{(0)}$. 
Since it comes from the linear terms 
for $h_{(0)}$ in Eq.(\ref{on-shell}), 
the result involves the integrals of 
$\ep^2(y)\dlt(y-\vth R)$ or $\dlt^2(y-\vth R)$.  
These quantities must be evaluated under some regularization of 
the orbifold singularity. 
If we pick up only the regularization-independent part of the result, 
we obtain 
\be
 m_h^2 = \frac{l^2}{6}m^2\brkt{1-\frac{2m}{k}}e^{-2k\pi R}
 \frac{1-e^{-2k\pi R}}{1-e^{-(k-2m)\pi R}}+\cO(l^4). 
 \label{mh}
\ee

Next, let us calculate the mass of $\chi_{h(0)}$, 
the (pseudo-) zero mode of $\chi_h$. 
For this purpose, we have to start from the full SUGRA action~\cite{KO2} 
because the hyperino~$\chi_h$ has mixing terms with the gravitino 
that we have dropped in the superspace action~(\ref{MO_action1}). 
In this calculation, we do not encounter the auxiliary fields 
in Eq.(\ref{on-shell}) and the result is independent of 
the orbifold regularization. 
The resulting mass term in the 4D effective Lagrangian is 
\be
 \cL^{(4)}_{\rm fermi} = i\bar{\chi}\gm^{\udl{m}}\der_m\chi
 -m_\chi\bar{\chi}\chi+\cdots, 
 \label{L_fermi}
\ee
where $\chi$ is the following Dirac spinor 
\be
 \chi \equiv \vct{\chi_T}{\bar{\chi}_{h(0)}}. \label{def_chi}
\ee
Here, $\chi_T$ is the lightest mode of 
$\psi_y^-\equiv i(\psi_{y{\rm R}}^{i=2}+\psi_{y{\rm L}}^{i=1})$, 
which corresponds to the superpartner of the radion. 
The mass~$m_\chi$ is 
\be
 m_\chi = \frac{l}{\sqrt{6}}\brkt{\frac{3}{2}k+m}
 \brkt{1-\frac{2m}{k}}^{1/2}e^{-k\pi R}\brkt{
 \frac{1-e^{-2k\pi R}}{1-e^{-(k-2m)\pi R}}}^{1/2}+\cO(l^3). 
 \label{m_chi}
\ee
Since $\chi_{h(0)}$ has a Dirac mass with $\chi_T$, 
these two fermionic modes are degenerate. 
Note that $m_\chi$ in Eq.(\ref{m_chi}) equals 
the radion mass~$m_{\rm rad}$ calculated in Ref.~\cite{AS2}. 
This is a trivial result because $\chi_T$ is a superpartner of the radion. 
Therefore, we can conclude that the correct value of 
the mass of $h_{(0)}$ is $m_{\rm rad}$ due to the $\cN=1$ supersymmetry 
preserved by the background~(\ref{bkgd}). 
Comparing Eq.(\ref{mh}) to Eq.(\ref{m_chi}), 
we can see that terms proportional to $km$ and $k^2$ are missing 
in Eq.(\ref{mh}).  
These missing terms are expected to be provided from terms involving 
$\ep^2(y)\dlt(y-\vth R)$ and $\dlt^2(y-\vth R)$, respectively. 
Therefore, a regularization scheme that respects 
the preserved supersymmetry is necessary 
to calculate $m_h^2$ correctly. 
This is a generic problem in the 5D off-shell approach 
if we work in a model with boundary terms. 
Especially, in the case that the background is non-BPS, 
we cannot estimate quantities involving the 
regularization-dependent terms without the information about 
the more fundamental theory, such as the string theory.

\subsection{Comment on Marti-Pomarol action} \label{cmt_MP}
Before ending this section, we will briefly comment on the relation 
to the superspace action in Ref.~\cite{MP}. 
In the 5D off-shell approach, there are four {\it dependent} 
superfields~($\Xi,\Xi^c,V^0,E$)~\footnote{
Their scalar and spinor components are rewritten 
by the other physical fields. 
See Eq.(\ref{GF_cond}). }
and one $\cN=1$ vector superfield~$V_E$ which originates from 
the fifth-dimensional component of the 5D gravitational multiplet. 
All the other superfields are the ordinary matter superfields. 
In fact, the five superfields 
\be
 \Xi,\;\; \Xi^c,\;\; V^0,\;\; E,\;\; V_E  \label{Sgm-like}
\ee  
characterize the theory as the supergravity. 
Besides their auxiliary fields and the graviphoton field~$W_\mu^0$, 
all the components of the five multiplets become constant 
at the leading order of $\kp$. 
Beyond the leading order, they induce all the SUGRA interactions 
which are generically suppressed by $\kp$. 
In this sense, they play a similar role to 
the chiral compensator superfield~$\Sgm$ 
in 4D off-shell SUGRA~\cite{KU}. 
So we will refer to the superfields~(\ref{Sgm-like}) 
as ``$\Sgm$-like superfields" here. 

If we neglect $\kp$-suppressed quartic or higher couplings, 
the fermionic components of the $\Sgm$-like superfields 
can be dropped. 
Roughly speaking, the equation of motion for $D^0$ is used to 
determine the background geometry, \ie, $\sgm(y)$. 
After the geometry is determined, $V^0$ and $\Xi^c$ contribute 
to only $\kp$-suppressed terms we are neglecting here. 
Furthermore, if the bulk scalar fields do not have any nontrivial 
background configurations, $V_E$ is related to $E$ as~\cite{AS} 
\be
 V_E \simeq \frac{E+\bar{E}}{2}, \label{VE_E}
\ee
up to the $\kp$-suppressed terms. 
This can be seen from the second equation in Eq.(\ref{aux_EOM}). 
The second term in the right-hand side of that equation leads 
to only $\kp$-suppressed quartic terms if $h_{\rm cl}=h^c_{\rm cl}=0$. 
Thus, $\cF_E \simeq 2\cF_V$. 
This suggests Eq.(\ref{VE_E}). 

Therefore, there remain only two $\Sgm$-like 
superfields~$\Xi$ and $E$ in this case. 
They correspond to the superfields called 
as the compensator and the radion superfields in Ref.~\cite{MP}. 
Using the relation~(\ref{VE_E}), our superspace action becomes 
a similar form to that of Ref.~\cite{MP}. 
As we have pointed out in Ref.~\cite{AS}, however, 
there is a difference between them that cannot be removed 
by the superfield redefinition. 
Namely, the action in Ref.~\cite{MP} is not consistent 
with the 5D SUGRA action of Refs.~\cite{KO,KO2}. 
Furthermore, we should treat $\cF_V$ and $\cF_E$ 
as independent auxiliary fields in the case that 
the bulk scalars have nonzero background configurations, 
as in the case of Sect.~\ref{SUSY_stblz}. 
In fact, the relation $\cF_E \simeq 2\cF_V$ does not hold 
in Eq.(\ref{on-shell}). 

Of course, all five $\Sgm$-like superfields are necessary 
for realizing the full SUGRA action 
including the $\kp$-suppressed quartic or higher terms.

\section{4D off-shell approach} \label{4Doffshell}
In this section, we will explain another approach to derive 
the 4D effective theory in the case that the background preserves 
$\cN=1$ SUSY. 
In this case, the effective theory is expected to be 4D SUGRA. 
Thus it can be expressed within the framework of 
4D off-shell SUGRA~\cite{KU}. 
In this framework, there is only one superfield 
other than the ordinary matter superfields, \ie, 
a chiral compensator superfield~$\Sgm$. 
This $\Sgm$ is not obtained from the 5D compensator superfields 
by the usual K.K. decomposition. 
Since the 5D compensator superfields are not independent 
of the matter fields, the usual K.K. decomposition cannot be performed 
for them. 
Therefore, we need a different approach 
to obtain the 4D off-shell action. 
The key of this approach is the introduction of the radion superfield. 
The resultant action becomes a 4D superspace action. 
Thus we will refer to this approach as the 4D off-shell approach in this paper. 

In order to obtain the complete effective action, we have to take 
all the gravitational interactions into account. 
In this section, however, we will focus on the matter-radion couplings 
among them since the other gravitational interactions 
are generically suppressed by the large 4D Planck mass 
and are irrelevant to the phenomenological discussion 
in many realistic brane-world models. 
The matter-radion couplings, on the other hand, cannot be neglected 
in such models because they play an important role in the mediation 
of SUSY breaking effects when we introduce 
some SUSY breaking mechanism.\footnote{
The 4D off-shell approach remains applicable even in such a case 
if the SUSY breaking scale is much lower than 
the Kaluza-Klein mass scale, which is a cutoff scale of 
the 4D effective theory.} 
Namely the $F$-term of the radion superfield mediates 
the SUSY breaking effects to the visible sector through 
the matter-radion couplings. 

\subsection{Radion superfield in 5D action}
We should note that the physical matter multiplets 
can be expanded into K.K. towers of 4D multiplets in the ordinary manner 
at least the leading order for the $\kp$-expansion. 
For a 5D matter superfield~$\Phi$, 
the mode expansion can be performed 
keeping the $\cN=1$ superfield structure as 
\be
 \Phi(x,y,\tht) = \sum_{n=0}^\infty f_{(n)}(y)\vph_{(n)}(x,\tht), 
 \label{SUSY_md_ex}
\ee
where $f_{(n)}(y)$ are appropriate mode functions that are 
solutions of the mode equations.  
Since the only obstacle for the naive K.K. expansion is 
the existence of the compensator and graviphoton multiplets, 
we can derive the 4D superspace action just like the global SUSY case 
if we drop them. 
Of course, the action derived in such a way lacks the radion superfield 
not only the 4D gravitational interaction terms. 
Although the radion multiplet behaves as a chiral matter multiplet 
in the 4D effective theory, 
its dependence of the effective action is not determined 
by the ordinary K.K. dimensional reduction 
in contrast to the other matter multiplets.
This stems from the fact that the radion originally belongs to 
the 5D gravitational field. 

In our previous paper~\cite{AS2}, we have clarified the appearance of 
the radion superfield~$T$ in the 5D superspace action as follows.\footnote{ 
In the following, we will neglect the backreaction on the metric, 
\ie, $\sgm(y)=-k\abs{y}$. } 
\bea
 \cL \eql \cL_{\rm bulk}
 +\sum_{\vth=0,\pi}\cL_{\rm brane}^{(\vth)}\dlt(y-\vth R), 
 \nonumber\\
 \cL_{\rm bulk} \eql \brc{\int\dr^2\tht\;G(T)\cW^2+\hc}
 +e^{2\sgm}\int\dr^4\tht\;
 \GR^{-2}(T)\brkt{\der_y V+i\Phi_S-i\bar{\Phi}_S}^2 \nonumber\\
 &&+e^{2\sgm}\int\dr^4\tht\;\brc{\GR^{\frac{3}{2}}(T)
 \brkt{\bar{H}e^{2gV}H+\bar{H}^ce^{-2gV}H^c}-3M_5^3\ln\GR} 
 \nonumber\\
 &&+e^{3\sgm}\brc{\int\dr^2\tht\;
 H^c\brkt{\der_y+\frac{3}{2}\dot{\sgm}+mG(T)-2ig\Phi_S}H+\hc}, 
 \nonumber\\
 \cL_{\rm brane}^{(\vth)} \eql \brc{\int\dr^2\tht\;
 f^{(\vth)}_{AB}(S)\cW^A\cW^B+\hc}
 -e^{2\sgm}M_5^2\int\dr^4\tht\;\GR^{-1}(T)
 \exp\brc{-K^{(\vth)}(S,\bar{S},U)} \nonumber\\
 &&+e^{3\sgm}\brc{\int\dr^2\tht\;G^{-\frac{3}{2}}(T)P^{(\vth)}(S)+\hc}, 
 \label{rad_action1}
\eea
where 
\bea
 G(T) \defa \brkt{1+e^{2k\abs{y}}e^{-k\pi R}
 \frac{\sinh k\pi(R-T)}{\sinh k\pi T}}^{-1}, \nonumber\\
 \GR(T) \defa \Re G(T). \label{def_G}
\eea
The 4D vector and chiral superfields~$U$ and $S$ 
can be localized or induced superfields on the boundaries. 
In the case that $S$ is an induced superfield 
from a bulk hypermultiplet~$(H,H^c)$, it is understood as~\footnote{
$H^c$ is $Z_2$-odd and cannot appear in the boundary actions. } 
\be
 S = G^{\frac{3}{4}}(T)H.  \label{def_S}
\ee
The power of $G(T)$ in Eq.(\ref{rad_action1}) is determined 
by the Weyl-weight matching and 
the condition for the radion mode to be promoted to a chiral superfield. 
(See Sect.3.2 in Ref.~\cite{AS2}.) 
The last term in the second line of $\cL_{\rm bulk}$ corresponds to 
the kinetic term for the radion superfield. 
Roughly speaking, the scalar component of $G(T)$ corresponds 
to $\ge{y}{4}$.\footnote{
Here, we have chosen the coordinate~$y$ so that $\gey=1$. 
Thus, the parameter~$R$ is the radius of the orbifold. 
Namely, $\vev{T}=R$ by definition of $T$. } 
The radion dependence of the metric for our result 
agrees with that of Ref.~\cite{BNZ}, 
where the $T$-dependent effective action for 5D pure SUGRA is derived. 

Note that Eq.(\ref{rad_action1}) should not be understood as a 5D action. 
This is because $T$ is introduced as a 4D superfield from the beginning 
in this approach. 
We cannot eliminate the auxiliary field of $T$ 
at this stage because it has no $y$-dependence. 
Thus, Eq.(\ref{rad_action1}) is not a final form yet. 
Recall that the K.K. expansion can be performed in the ordinary way 
for the other matter superfields.  
In order to obtain the 4D effective action, 
we have to expand them into K.K. modes first, drop the heavier modes 
than the cut-off scale of the effective theory, 
and perform the $y$-integral. 
Then we can obtain the desired effective action. 

Let us demonstrate this procedure in the model of Ref.~\cite{MO}. 
The hypermultiplet~$(H,H^c)$ is expanded as 
\bea
 H(x,y,\tht) \eql \sum_{n=0}^\infty f_{(n)}(y)H_{(n)}(x,\tht), \nonumber\\
 H^c(x,y,\tht) \eql h_{\rm cl}^c(y)
 +\sum_{n=1}^\infty f^c_{(n)}(y)H^c_{(n)}(x,\tht), 
 \label{hyp_md_ex}
\eea
where $h_{\rm cl}^c(y)$ is defined in Eq.(\ref{bkgd}). 
Here we will neglect the subleading contribution for $l$-expansion. 
In this case, $H^c$ has no light mode, 
while $H$ has a (pseudo) zero mode whose mode function is 
\be
 f_{(0)}(y) = C_he^{\brkt{\frac{3}{2}k-m}\abs{y}}, 
\ee
where $C_h$ is defined in Eq.(\ref{def_Ch}). 
The mode functions of the massive modes for $H$ and $H^c$ 
are expressed by the Bessel functions~\cite{GP}. 
(See Appendix~\ref{mf_HHc}.)

The function~$G(T)$ is expanded as 
\be
 G(T) = 1+\alp e^{2k\abs{y}}\tl{T}+\cO(\tl{T}^2), \label{G_expd}
\ee
where $\tl{T}\equiv T-R$ is a fluctuation of the radion superfield 
around the background value, and 
\be
 \alp\equiv\frac{2k\pi}{e^{2k\pi R}-1}. 
\ee

Plugging these expressions into Eq.(\ref{rad_action1})  
and performing the $y$-integral, we obtain the effective action, 
\bea
 \cL^{(4)} \eql \int\dr^4\tht\;\brc{\frac{3}{4}M_5^3\pi\alp\abs{\tl{T}}^2
 +\abs{H_{(0)}}^2+\cO(\tl{T}^3)} \nonumber\\
 &&+\sbk{\int\dr^2\tht\;\brc{\frac{J_0C_h\pi}{2}\brkt{\frac{3}{2}k+m}
 \tl{T}H_{(0)}+\cO(\tl{T}^2)}+\hc}+\cdots, 
 \label{4D_L1}
\eea
where the ellipsis denotes terms involving the massive modes.

From Eq.(\ref{4D_L1}), we can easily see that 
$\tl{T}$ and $H_{(0)}$ has the following degenerate mass. 
\be
 m_{\rm rad}=m_{H(0)}=\frac{l}{\sqrt{6}}\brkt{\frac{3}{2}k+m}
 \brkt{1-\frac{2m}{k}}^{1/2}e^{-k\pi R}\brkt{
 \frac{1-e^{-2k\pi R}}{1-e^{-(k-2m)\pi R}}}^{1/2}. 
 \label{m_rad}
\ee
This certainly agrees with the result~(\ref{m_chi}) 
in the previous section. 
Note that $\chi_T$ and $\chi_{h(0)}$ in Eq.(\ref{def_chi}) 
are the fermionic components of $\tl{T}$ and $H_{(0)}$, respectively. 

In contrast to the 5D off-shell approach, 
we have not encountered any regularization-dependent quantities 
such as $\ep^2(y)\dlt(y-\vth R)$ or $\dlt^2(y-\vth R)$. 
This is because we have not eliminated the auxiliary fields 
until we derive the 4D superspace action~(\ref{4D_L1}). 
Therefore, we do not have to search for a supersymmetric regularization 
of the orbifold singularity. 
This is one of the advantage of the 4D off-shell approach. 

In general, the radion K\"{a}hler potential~$K_{\rm rad}^{(4)}(T,\bar{T})$ becomes 
a complicated function because of the nontrivial $y$ dependence of $G(T)$ 
in Eq.(\ref{def_G}). 
However, in the flat limit (\ie, $k\to 0$), $G(T)$ becomes independent of $y$ 
and reduces to a simple form, 
\be
 G(T) =\frac{T}{R}=1+\frac{\tl{T}}{R}. 
\ee
In this case, the radion K\"{a}hler potential has the following no-scale form 
up to a constant. 
\be
 K_{\rm rad}^{(4)}(T,\bar{T})=-3M_4^2\ln(T+\bar{T}), 
\ee
where $M_4\equiv (\pi RM_5^3)^{1/2}$ is the 4D Planck mass.

\subsection{Consistency with 5D off-shell approach} \label{consistency}
Now we will check the consistency of this 4D off-shell approach 
with the 5D off-shell approach discussed in the previous section. 
The agreement of Eqs.(\ref{m_chi}) and (\ref{m_rad}) provides a nontrivial 
cross-check for the powers of $G(T)$ in the boundary superpotential 
and in Eq.(\ref{def_S}). 
For further checks, we will introduce 
an extra hypermultiplet~$(Q,Q^c)$ with a bulk mass~$m_Q$ 
and a vector multiplet~$(V,\Phi_S)$ to the model of Ref.~\cite{MO}. 
The mode expansions of these additional superfields are as follows. 
\bea
 Q(x,y,\tht) \eql \sum_{n=0}^\infty f_{Q(n)}(y)Q_{(n)}(x,\tht), \nonumber\\
 Q^c(x,y,\tht) \eql \sum_{n=1}^\infty f_{Q(n)}^c(y)Q^c_{(n)}(x,\tht), \nonumber\\
 V(x,y,\tht) \eql \sum_{n=0}^\infty v_{(n)}(y)V_{(n)}(x,\tht), \nonumber\\
 \Phi_S(x,y,\tht) \eql \sum_{n=1}^\infty u_{(n)}(y)\Phi_{S(n)}(x,\tht). 
\eea
Note that $Q^c$ and $\Phi_S$ do not have zero-modes 
because they are odd under $Z_2$-parity. 
The explicit forms of the mode functions are collected in Appendix~\ref{mf_HHc}. 

In 5D off-shell SUGRA, 
there are five superfields~(\ref{Sgm-like}) that play the similar role 
to that of the chiral compensator~$\Sgm$ in 4D off-shell SUGRA, 
as mentioned in Sect.~\ref{cmt_MP}. 
They induce all the gravitational interactions, which partly become 
the couplings to the radion superfield. 
Thus, to check the consistency between the two approaches, 
we will focus on the contributions from the superfields~(\ref{Sgm-like}) 
in the 5D off-shell approach and those from $T$ in the 4D off-shell approach 
because the other parts are identical in both approaches. 

In the 5D off-shell approach, we can see from Eq.(\ref{on-shell}) 
that $\cF_E$, $\cF_V$ and $\cF_\xi$ 
have linear terms for $h_{(0)}$ in their on-shell values. 
There are singular terms among them that are proportional to 
$\dlt(y-\vth R)$ ($\vth=0,\pi$). 
They lead to the regularization-dependent terms. 
Note that the non-singular terms in Eq.(\ref{on-shell}) are proportional to 
the bulk mass parameter~$m$ while the singular terms do not. 
This means that the $m$-dependent part of each quantity is completely calculated 
only from the non-singular terms. 
Thus we will focus on the $m$-dependent contributions here. 
The linear terms in Eq.(\ref{on-shell}) lead to the following cubic interactions.  
\bea
 \cL_{\rm cubic}^{(4)} \eql \frac{\kp^3mJ_0C_h}{6}\left\{
  \sum_{n,l}\brkt{a^\lmd_{nl}\bar{h}_{(0)}\lmd_{(n)}\lmd_{(l)}
  +2a^\chi_{nl}h_{(0)}\chi_{S(n)}\chi_{S(l)}} \right. \nonumber\\
  &&\left.\hspace{21mm}+\sum_{n,l}3b_{nl} \bar{h}_{(0)}q_{(n)}^c q_{(l)}
  +\hc\right\}+\cdots, 
 \label{L_cubic1}
\eea
where $\lmd_{(n)}$, $\chi_{S(n)}$ and $q_{(l)}$, $q^c_{(l)}$ 
are the spinor components of $V_{(n)}$, $\Phi_{S(n)}$ 
and the scalar components of $Q_{(l)}$ and $Q^c_{(l)}$, respectively. 
The ellipsis denotes the $m$-independent contributions 
coming from the singular terms in Eq.(\ref{on-shell}). 
The constants~$a^\lmd_{nl}$, $a^\chi_{nl}$ and $b_{nl}$ are defined as 
\bea
 a^\lmd_{nl} \defa \int_0^{\pi R}\dr y\;
 e^{2ky}v_{(n)}(y)v_{(l)}(y), \nonumber\\
 a^\chi_{nl} \defa \int_0^{\pi R}\dr y\;
 u_{(n)}(y)u_{(l)}(y), \nonumber\\
 b_{nl} \defa \brc{m^q_{(n)}d_{nl}+m^q_{(l)}d_{nl}^c
 -\frac{4m_Q}{3}c_{nl}},  
 \label{def_b}
\eea
where $m^q_{(n)}$ are degenerate mass eigenvalues of $(Q_{(n)},Q^c_{(n)})$, 
and 
\bea
 c_{nl} \defa \int_0^{\pi R}\dr y\; e^{-ky}f^c_{Q(n)}(y)f_{Q(l)}(y), \nonumber\\
 d_{nl} \defa \int_0^{\pi R}\dr y\; f_{Q(n)}(y)f_{Q(l)}(y), \nonumber\\
 d^c_{nl} \defa \int_0^{\pi R}\dr y\; f^c_{Q(n)}(y)f^c_{Q(l)}(y). 
\eea

In the 4D off-shell approach, 
the corresponding interaction terms come from 
the contribution of $\cF_T$, the auxiliary field of $T$. 
The 4D effective Lagrangian is calculated as follows by repeating 
the procedure from Eq.(\ref{rad_action1}) to Eq.(\ref{4D_L1}) 
including $(Q,Q^c)$ and $(V,\Phi_S)$. 
The result is 
\bea
 \cL^{(4)} \eql \frac{1}{4}\brc{\int\dr^2\tht\;\sum_n \cW_{(n)}^2
 +\alp\sum_{n,l}a^\lmd_{nl}\tl{T}\cW_{(n)}\cW_{(l)}+\hc} \nonumber\\
 &&+\int\dr^4\tht\;\sbk{\sum_{n\neq 0}\abs{\Phi_{S(n)}}^2
 +\alp\sum_{n,l}\brc{a^\chi_{nl}\tl{T}\bar{\Phi}_{S(n)}\bar{\Phi}_{S(l)}+\hc}} 
 \nonumber\\
 &&+\int\dr^4\tht\;
 \left[\frac{3}{4}M_5^3\pi\alp\abs{\tl{T}}^2
 +\abs{H_{(0)}}^2+\sum_n\abs{Q_{(n)}}^2
 +\sum_{n\neq 0}\abs{Q^c_{(n)}}^2 \right. \nonumber\\
 &&\hspace{25mm}\left.
 +\frac{3}{2}\alp\sum_{n,l}\Re\tl{T}\brc{d_{nl}\bar{Q}_{(n)}Q_{(l)}
 +d^c_{nl}\bar{Q}^c_{(n)}Q^c_{(l)}}\right] \nonumber\\
 &&+\left[\int\dr^2\tht\;\left\{
 \frac{\pi J_0C_h}{2}\brkt{\frac{3}{2}k+m}\tl{T}H_{(0)}
 +\sum_{n\neq 0}m^q_{(n)}Q^c_{(n)}Q_{(n)} \right.\right. \nonumber\\
 &&\hspace{25mm}\left.\left.+m_Q\alp\sum_{n,l}c_{nl}
 \tl{T}Q^c_{(n)}Q_{(l)}\right\}+\hc\right]+\cdots, \label{4D_L3}
\eea
where the ellipsis denotes irrelevant terms to the discussion here. 

From this action, the on-shell value of $\cF_T$ is calculated as
\be
 \cF_T = \frac{2\kp^3J_0 C_h}{3\alp}\brkt{\frac{3}{2}k+m}\bar{h}_{(0)}+\cdots, 
 \label{F_T}
\ee
where the ellipsis denotes terms beyond the linear order for fields 
or involving the massive modes~$h_{(n)}$ ($n\neq 0$). 
The linear term for $h_{(0)}$ in Eq.(\ref{F_T}) induces 
the following cubic couplings. 
\bea
 \cL_{\rm cubic}^{(4)} \eql \frac{\kp^3J_0C_h}{6}\brkt{\frac{3}{2}k+m}\left\{
 \sum_{n,l}\brkt{a^\lmd_{nl}\bar{h}_{(0)}\lmd_{(n)}\lmd_{(l)}
 +2a^\chi_{nl}h_{(0)}\chi_{S(n)}\chi_{S(l)}} \right. \nonumber\\
 &&\left.\hspace{38mm}+\sum_{n,l}3b_{nl}\bar{h}_{(0)}q^c_{(n)}q_{(l)}+\hc\right\}.  
 \label{L_cubic2}
\eea
This result is consistent with that obtained in 
the 5D off-shell approach~(\ref{L_cubic1}). 
Note that we have encountered no regularization-dependent terms. 

The matching of the masses and the cubic interactions 
obtained in both approaches provides nontrivial cross-checks 
for their consistency. 
Note that these mass and cubic terms come from 
the linear terms for $h_{(0)}$ in $\cF_E$, $\cF_V$ and $\cF_\xi$ 
in the 5D off-shell approach (see Eq.(\ref{on-shell})), 
while they come from the linear term in $\cF_T$ 
in the 4D off-shell approach (see Eq.(\ref{F_T})). 
This means that the role of the superfields~(\ref{Sgm-like}) 
in the 5D off-shell approach is partially played by 
the radion superfield~$T$ in the 4D off-shell approach. 


Finally, we will comment on the result of Ref.~\cite{LS}, in which  
the authors constructed the 4D off-shell effective action of 5D SUGRA. 
However, their method is based on the naive replacement of the radius of 
the orbifold~$R$ with the dynamical radion field~$r(x)$. 
We have pointed out in our previous work~\cite{AS2,AS3} 
that this naive replacement does not lead to the correct action. 
This is because the mode function of the radion is not taken into account 
by this replacement. 
In fact, our result obtained in the 4D off-shell approach 
does not agree with the result of Ref.~\cite{LS}.

\section{Summary} \label{summary}
We have discussed the dimensional reduction of 5D off-shell SUGRA 
to derive the 4D effective theory. 
There are two approaches to derive the 4D effective action. 
We call them the 5D and the 4D off-shell approaches in this paper. 
The essential difference between them is the treatment of the compensator 
and the radion superfields in the action. 

The 5D off-shell approach is based on the 5D superconformal gravity 
formulated in Refs.~\cite{KO,KO2,FKO}. 
Since each 5D superconformal multiplet can be decomposed into 
$\cN=1$ multiplets~\cite{KO}, we can express the 5D SUGRA action 
in the language of 4D off-shell SUGRA~\cite{PST,AS}. 
Therefore, it seems possible to treat 5D off-shell SUGRA 
as if 4D off-shell SUGRA at first sight. 
However, there are five $\cN=1$ multiplets~(\ref{Sgm-like}) 
that have no counterpart in 4D off-shell SUGRA. 
These five $\cN=1$ multiplets play a similar role 
to the chiral compensator multiplet~$\Sgm$ in 4D off-shell SUGRA. 
If the bulk scalars do not have any nontrivial background configurations, 
only $\Xi$ and $E$ among them remain in the action 
if we neglect $\kp$-suppressed quartic or higher terms. 
(See Sect.\ref{cmt_MP}.) 
This corresponds to the situation considered in Ref.~\cite{MP}. 
There, $\Xi$ and $E$ are called as 
the compensator and the radion superfields, respectively. 
As we have pointed out in Ref.~\cite{AS}, however, 
their dependences of the action in Ref.~\cite{MP} 
is not consistent with the 5D SUGRA action of Refs.~\cite{KO,KO2}. 
Note also that the relation~(\ref{VE_E}) does not hold 
if the bulk scalars have a nontrivial background configuration. 
For example, $\cF_V$ and $\cF_E$ should be treated as 
independent auxiliary fields in the model of Ref.~\cite{MO} 
discussed in Sect.\ref{SUSY_stblz}. 
Therefore, we should start from the action~(\ref{MO_action1}), 
which includes all the five superfields~(\ref{Sgm-like}). 

The important point in the 5D off-shell approach is that 
all the auxiliary fields must be eliminated 
{\it before} the dimensional reduction. 
This is because we cannot perform the naive K.K. expansion 
for the five superfields~(\ref{Sgm-like}). 
Thus the 4D effective action derived in this approach 
inevitably becomes the on-shell action. 
The on-shell values of the auxiliary fields generically have 
singular terms proportional to $\dlt(y-\vth R)$ ($\vth=0,\pi$) 
if the boundary terms exist. 
On the other hand, some quantities in the bulk are odd 
under the $Z_2$ parity. 
Therefore, some terms in the on-shell 5D action 
depend on the regularization of the orbifold singularity. 
This means that there are some quantities we cannot calculate 
without the orbifold regularization. 
This is one of the drawbacks of the 5D off-shell approach. 

In the case that the background preserves $\cN=1$ SUSY, 
there is another approach to derive the 4D effective action, 
which we call the 4D off-shell approach.  
The key of this approach is an introduction of the radion superfield~$T$. 
The starting point is Eq.(\ref{rad_action1}), 
which shows the $T$-dependence of the 5D action. 
Note that all 5D $\cN=1$ superfields in Eq.(\ref{rad_action1}) 
can be expanded into K.K. modes 
keeping the structure of $\cN=1$ superfields. 
Performing the $y$-integral, the 4D effective action written 
in the superspace is obtained. 
Although Eq.(\ref{rad_action1}) is not the final form of the above derivation, 
it is useful because we can easily read off integrands of 
the overlap integrals that appear as coupling constants 
in the effective action. 
Since we obtain the 4D effective action as a superspace action, 
the preserved $\cN=1$ SUSY is manifest. 
In addition, we do not suffer from the regularization-dependent terms 
in contrast to the 5D off-shell approach. 
Recall that all such terms originate from the elimination of 
the 5D auxiliary fields. 
In the present case, on the other hand, 
the elimination of the auxiliary fields does not 
cause such terms because the $y$-dependence is already integrated out. 
We have focus on the radion couplings among the gravitational interactions 
in this approach. 
In order to incorporate all the interactions into the discussion, 
we have to find some method to derive Eq.(\ref{rad_action1}) directly from 
the original 5D off-shell SUGRA action. 
We will discuss this issue in the forthcoming paper. 

We have checked the consistency between the effective actions 
derived by the above two approaches. 
Comparing Eqs.(\ref{MO_action1}) and (\ref{rad_action1}), 
the difference between the two approaches is the treatment of 
the compensator and the radion superfields. 
The role of the five superfields~(\ref{Sgm-like}) in the 5D off-shell approach 
is partially inherited by $T$ in the 4D off-shell approach. 

When we use the word~``the radion superfield", 
we have to specify the context in which we work. 
It denotes $E$ (and $V_E$) in the 5D off-shell approach,  
while it denotes $T$ in the 4D off-shell approach. 
For example, Ref.~\cite{MP} corresponds to the former case, 
and Ref.~\cite{LS} to the latter case although their actions should be modified. 
Here note that only the 4D off-shell approach can treat it 
as a dynamical superfield. 
In the 5D off-shell approach, it becomes a spurion-like superfield 
since the gravitational fields are fixed to the background values 
in compensation for expressing the action on the superspace.\footnote{
If we use the $F$- and $D$-term formulae instead of the superspace, 
we can treat ``the radion multiplet" $E$ (and $V_E$) as dynamical fields 
also in the 5D off-shell approach, 
but the action becomes quite complicated. }

One of the advantage of the 5D off-shell approach is 
that it can also deal with non-BPS backgrounds.  
Especially, the Scherk-Schwarz SUSY breaking mechanism~\cite{SS} is 
clearly understood in this approach~\cite{AS,AS4}. 
Its disadvantage is the appearance of the regularization-dependent 
quantities. 
The regularization of the orbifold depends on the fundamental theory 
at higher energies. 
Namely, such quantities are UV sensitive and cannot be calculated 
without specifying the UV theory.  
On the other hand, the 4D off-shell approach can derive 
the effective action without any regularization-dependent 
quantities with the aid of the radion superfield. 
In this approach, the preserved SUSY is manifest 
since the effective action is expressed by superfields. 
However, it should be noted that the 4D off-shell approach is available 
only when the background is BPS. 
For example, it cannot deal with the Scherk-Schwarz SUSY breaking. 

As a future work, we will plan to extend the discussion to 
the case where the bulk is six or higher dimensions. 
Unfortunately, it is known that no off-shell description of 
such higher dimensional SUGRA with matter multiplets exists. 
Therefore, only the 4D off-shell approach is available. 
In such a case, more moduli superfields will appear 
and the structure of the moduli space becomes rich and complicated. 
Thus, the investigation of such structure is also  an intriguing subject. 

\vspace{3mm}

\begin{center}
{\bf Acknowledgments}
\end{center}
This work was supported by the Japan Society for the Promotion of Science 
for Young Scientists (No.0509241) (Y.S.).

\appendix

\section{Superfields and gauge fixing} \label{SF_GF}
Here, we will collect the explicit forms of the $\cN=1$ superfields 
in terms of the superconformal notation of Refs.~\cite{KO,KO2}. 

\subsection{4D superfields}
4D superconformal multiplets can appear in the boundary action. 
The construction of $\cN=1$ superfields from them is straightforward. 
The only thing to note is the existence of the warp factor. 
This comes from the $y$-dependence of the Killing spinor 
for the preserved SUSY. 

From a vector multiplet~$(B_m^A,\lmd^A,D^A)$ ($A=1,2,\cdots$), 
we can obtain the following vector superfield. 
\be
 U^A \equiv \tht\sgm^{\udl{m}}\btht B_m^A
 +ie^{\frac{3}{2}\sgm}\tht^2\btht\bar{\lmd}^A
 -ie^{\frac{3}{2}\sgm}\btht^2\tht\lmd^A
 +\frac{1}{2}e^{2\sgm}\tht^2\btht^2 D^A, 
\ee
where $\sgm(y)$ is estimated at the boundary~$y=0,\pi R$. 
The corresponding superfield strength is defined as 
\be
 \cW^A_\alp \equiv -\frac{1}{4}\bar{D}^2 D_\alp U^A. 
\ee

From a chiral multiplet~$(s^a,\chi_s^a,\cF_s^a)$ $(a=1,2,\cdots)$, 
we can obtain the following chiral superfield. 
\be
 S^a \equiv s^a-e^{\frac{\sgm}{2}}\tht\chi_s^a
 -e^\sgm\tht^2\cF_s^a. 
\ee

\subsection{5D superfields} \label{5Dsf}
From the 5D Weyl multiplet, 
we can construct the following $\cN=1$ superfield. 
\be
 V_E = \gey+ie^\sgm \tht^2\brkt{V_y^{(1)}+iV_y^{(2)}}
 -ie^\sgm \btht^2\brkt{V_y^{(1)}-iV_y^{(2)}}, 
\ee
where $V_y^{(r)}$ is an $\SUu$ gauge field, which is an auxiliary field 
in 5D off-shell SUGRA. 
This superfield corresponds to the $\cN=1$ multiplet~$\bdm{W}_y$ 
in Ref.~\cite{KO}. 

A vector multiplet~$\cV^I$ ($I=0,1,\cdots,\nV$) consists of 
\be
 M^I,\;\;\; W_\mu^I, \;\;\; \Omg^{Ii}, \;\;\; Y^{I(r)}, 
\ee
which are a gauge scalar, a gauge field, a gaugino 
and an auxiliary field, respectively. 
The indices $i=1,2$ and $r=1,2,3$ are doublet and triplet indices 
for $\SUu$. 
From this multiplet, we can construct the following $\cN=1$ vector 
and chiral superfields. 
\bea
 V^I \defa \tht\sgm^{\udl{m}}\btht W_m^I+i\tht^2\btht\bar{\lmd}^I
 -i\btht^2\tht\lmd^I+\frac{1}{2}\tht^2\btht^2 D^I, \nonumber\\
 \Phi_S^I \defa \vph_S^I-\tht\chi_S^I-\tht^2\cF_S^I, 
\eea
where
\bea
 \lmd^I \defa 2e^{\frac{3}{2}\sgm}\Omg_{\rm R}^{I1}, \nonumber\\
 D^I \defa -e^{2\sgm}\brc{\gey^{-1}\der_y M^I-2Y^{I(3)}
 +\gey^{-1}\dot{\sgm}M^I}, \nonumber\\
 \vph_S^I \defa \frac{1}{2}\brkt{W_y^I+i\gey M^I}, \nonumber\\
 \chi_S^I \defa -2e^{\frac{\sgm}{2}}\gey\Omg_{\rm R}^{I2}, \nonumber\\
 \cF_S^I \defa e^\sgm\brc{\brkt{V_y^{(1)}+iV_y^{(2)}}M^I
 -i\gey\brkt{Y^{I(1)}+iY^{I(2)}}}. 
\eea

The hypermultiplets consist of complex scalars~$\cA_i^\alp$, 
spinors~$\zeta^\alp$ and auxiliary fields~$\cF_i^\alp$. 
They carry a $\Usp$ index~$\alp=1,2,\cdots,2\nH+2$ 
on which the gauge group can act. 
These are split into $\nH+1$ hypermultiplets as 
\be
 \cH^{\hat{\alp}}=\brkt{\cA^{2\hat{\alp}+1}_i, \cA^{2\hat{\alp}+2}_i,
 \zeta^{2\hat{\alp}+1}, \zeta^{2\hat{\alp}+2}, \cF^{2\hat{\alp}+1}_i, 
 \cF^{2\hat{\alp}+2}_i}. \;\;\;\;\; 
 (\hat{\alp}=0,1,\cdots,\nH) 
\ee
From these multiplets, we can construct 
the following chiral superfields. 
\be
 \Phi^\alp \equiv \vph^\alp-\tht\chi^\alp-\tht^2\cF^\alp, 
\ee
where 
\bea
 \vph^\alp \defa \cA^\alp_2, \nonumber\\
 \chi^\alp \defa -2ie^{\frac{\sgm}{2}}\ztR^\alp,  \\
 \cF^\alp \defa e^\sgm\gey^{-1}\left\{\der_y\cA^\alp_1
 +i\brkt{V_y^{(1)}+iV_y^{(2)}}\cA_2^\alp
 +i\brkt{\gey+\frac{iW_y^0}{M^0}}\tl{\cF}_1^\alp \right. \nonumber\\
 &&\hspace{20mm}\left.-\brkt{W_y^I-i\gey M^I}(gt_I)^\alp_{\;\;\bt}\cA_1^\bt
 +\frac{3}{2}\dot{\sgm}\cA_1^\alp\right\}.  \nonumber
\eea

\subsection{Superconformal gauge-fixing conditions} \label{scGF}
The gauge fixing conditions for the extraneous superconformal 
symmetries, \ie, the dilatation~$\bdm{D}$, $\SUu$, 
the conformal supersymmetry~$\bdm{S}$ are as follows.\footnote{
The special conformal transformation~$\bdm{K}$ is already fixed 
in our superspace formalism~\cite{AS}. }

The $\bdm{D}$-gauge is fixed by 
\bea
 \cN \defa C_{IJK}M^I M^J M^K = M_5^3, \nonumber\\
 \cA_i^\alp\dmx\cA^i_\bt \eql 2\brc{-\sum_{\alp=1}^2\abs{\cA_2^\alp}^2
 +\sum_{\alp=3}^{2\nH+2}\abs{\cA_2^\alp}^2} = -2M_5^3, 
\eea
where $\cN$ is called the norm function. 

The $\SUu$ is fixed by 
\be
 \cA_i^\alp \propto \dlt_i^\alp. \;\;\;\;\; (\alp=1,2)
\ee

The $\bdm{S}$-gauge is fixed by 
\bea
 \cN_I\Omg^{Ii} \eql 0, \nonumber\\
 \cA_i^\alp\dmx\zeta_\bt \eql 0, 
\eea
where $\cN_I\equiv \der\cN/\der M^I$.

\section{Explicit forms of mode functions} \label{mf_HHc}
Here we will collect the explicit forms of the mode functions for 
the matter fields. 
On the Randall-Sundrum background~\cite{RS}, \ie, $\sgm(y)=-k\abs{y}$,
they are expressed by 
the Bessel functions~\cite{GP}. 

Let us define the following functions first. 
\bea
 \cM_\alp(\lmd) \defa J_\alp(\lmd)Y_\alp(\lmd e^{k\pi R})
 -Y_\alp(\lmd)J_\alp(\lmd e^{k\pi R}), \nonumber\\
 A_\alp(\lmd,z) \defa N_{\alp-1}(\lmd)
 \brc{Y_{\alp-1}(\lmd)J_\alp(\lmd z)-J_{\alp-1}(\lmd)Y_\alp(\lmd z)}, 
 \nonumber\\
 B_\alp(\lmd,z) \defa N_\alp(\lmd)
 \brc{Y_\alp(\lmd)J_\alp(\lmd z)-J_\alp(\lmd)Y_\alp(\lmd z)}, 
\eea
where 
\be
 N_\alp(\lmd) \equiv \frac{\sqrt{k}\pi\lmd}{\sqrt{2}}
 \brc{\frac{Y_\alp^2(\lmd)}{Y_\alp^2(\lmd e^{k\pi R})}-1}^{-1/2}. 
\ee

For the hypermultiplet~$(H,H^c)$, the mode expansions are as follows. 
\bea
 H(x,y,\tht) \eql \sum_{n=0}^\infty f_{(n)}(y)H_{(n)}(x,\tht), \nonumber\\
 H^c(x,y,\tht) \eql \sum_{n=1}^\infty f_{(n)}^c(y)H_{(n)}^c(x,\tht). 
\eea
For the zero mode~$H_{(0)}$, 
\be
 f_{(0)}(y) = C_h e^{\brkt{\frac{3}{2}k-m}\abs{y}}, \label{f0}
\ee
where the normalization constant~$C_h$ is 
\be
 C_h = \brkt{\frac{k-2m}{e^{(k-2m)\pi R}-1}}^{1/2}. \label{def_Ch}
\ee
For the massive modes~$H_{(n)}$, $H_{(n)}^c$ ($n\neq 0$),  
\bea
 f_{(n)}(y) \eql e^{2k\abs{y}}A_{c+\frac{1}{2}}(\lmd_n,e^{k\abs{y}}), 
 \nonumber\\
 f_{(n)}^c(y) \eql e^{2k\abs{y}}B_{c-\frac{1}{2}}(\lmd_n,e^{k\abs{y}})\ep(y), 
 \label{fn_fcn}
\eea
where $c\equiv m/k$, $\ep(y)$ is the periodic step function defined 
in Eq.(\ref{def_ep}), and $\lmd_n$ is a solution of 
\be
 \cM_{c-\frac{1}{2}}(\lmd_n) = 0. 
\ee
The corresponding mass eigenvalue is given by $m_{(n)}=k\lmd_n$. 
Note that $H_{(n)}$ and $H^c_{(n)}$ are degenerate for each nonzero $n$. 

For the vector multiplet~$(V,\Phi_S)$, 
the mode expansions are as follows. 
\bea
 V(x,y,\tht) \eql \sum_{n=0}^\infty v_{(n)}(y)V_{(n)}(x,\tht), \nonumber\\
 \Phi_S(x,y,\tht) \eql \sum_{n=1}^\infty u_{(n)}(y)\Phi_{S(n)}(x,\tht). 
\eea
For the zero mode~$V_{(0)}$, the mode function is 
\be
 v_{(0)}(y) = \frac{1}{\sqrt{\pi R}}. 
\ee
For the other massive modes~$V_{(n)}$, $\Phi_{S(n)}$ ($n\neq 0$), 
\bea
 v_{(n)}(y) \eql e^{k\abs{y}}A_1(\mu_n,e^{k\abs{y}}), \nonumber\\
 u_{(n)}(y) \eql \frac{e^{2k\abs{y}}}{\sqrt{2}}B_0(\mu_n,e^{k\abs{y}})\ep(y), 
\eea
where $\mu_n$ is a solution of 
\be
 \cM_0(\mu_n) = 0. 
\ee
The corresponding mass eigenvalue is given by $m_{(n)}=k\mu_n$. 
Note that $V_{(n)}$ and $\Phi_{S(n)}$ are degenerate for nonzero $n$. 

The above mode functions satisfy the following orthonormal relations. 
\bea
 \int_0^{\pi R}\dr y\; e^{-2ky}f_{(n)}(y)f_{(l)}(y) \eql \dlt_{nl}, 
 \;\;\;\;\;
 \int_0^{\pi R}\dr y\; e^{-2ky}f^c_{(n)}(y)f^c_{(l)}(y) = \dlt_{nl}, 
 \nonumber\\
 \int_0^{\pi R}\dr y\; v_{(n)}(y)v_{(l)}(y) \eql \dlt_{nl}, 
 \;\;\;\;\;
 \int_0^{\pi R}\dr y\; e^{-2ky}u_{(n)}(y)u_{(l)}(y) = \frac{\dlt_{nl}}{2}. 
\eea


\begin{thebibliography}{100}
 \bibitem{Cvetic} M.~Cveti\v{c}, H.~L\"{u} and C.N.~Pope, 
 {\it J.~Math.~Phys.}~{\bf 42} (2001) 3048 [{\tt hep-th/0009183}]. 

 \bibitem{Liu} J.T.Liu, H.~L\"{u} and C.N.~Pope, 
 {\tt hep-th/0212037}. 

 \bibitem{HW} P.~Ho\v{r}ava and E.~Witten, \NP{B460} (1996) 506 
 [{\tt hep-th/9510209}]; \NP{B475} (1996) 94 [{\tt hep-th/9603142}]. 

 \bibitem{RS} L.~Randall and R.~Sundrum, \PRL{83} (1999) 3370 
 [{\tt hep-ph/9905221}]. 

 \bibitem{GP} T.~Gherghetta and A.~Pomarol, \NP{B586} (2000) 141 
  [{\tt hep-ph/0003129}]; \NP{B602} (2001) 3 [{\tt hep-ph/0012378}].  

 \bibitem{ABN} R.~Altendorfer, J.~Bagger and D.~Nemeschansky, 
  \PR{D63} (2001) 125025 [{\tt hep-th/0003117}];
  A.~Falkowski, Z.~Lalak and S.~Pokorski, \PL{B491} (2000) 172 
  [{\tt hep-th/0004093}]. 

 \bibitem{BPS} E.B.~Bogomol'ny, {\it Sov.~J.~Nucl.~Phys.}~{\bf 24} (1976) 449;
  M.K.~Prasad and C.M.~Sommerfield, \PRL{35} (1975) 760. 

 \bibitem{KU} T.~Kugo and S.~Uehara, \NP{B226} (1983) 49;
  \PTP{73} (1985) 235. 

 \bibitem{KO} T.~Kugo and K.~Ohashi, \PTP{108} (2002) 203 
  [{\tt hep-th/0203276}]. 

 \bibitem{KO2} T.~Kugo and K.~Ohashi, \PTP{105} (2001) 323 
  [{\tt hep-h/0010288}]. 

 \bibitem{FKO} T.~Fujita, T.~Kugo and K.~Ohashi, \PTP{106} (2001) 671 
  [{\tt hep-th/0106051}]. 

 \bibitem{PST} F.~Paccetti Correia, M.G.~Schmidt and Z.~Tavartkiladze, 
  \NP{B709} (2005) 141 [{\tt hep-th/0408138}]. 

 \bibitem{AS} H.~Abe and Y.~Sakamura, \JH{0410} (2004) 013 
  [{\tt hep-th/0408224}]. 

 \bibitem{AGW} N.~Arkani-Hamed, T.~Gregoire and J.~Wacker, 
  \JH{0203} (2002) 055 [{\tt hep-th/0101233}];
  A.~Hebecker, \NP{B632} (2002) 101 [{\tt hep-ph/0112230}]. 

 \bibitem{MP} D.~Marti and A.~Pomarol, \PR{D64} (2001) 105025 
  [{\tt hep-th/0106256}]. 

 \bibitem{LS} M.A.~Luty and R.~Sundrum, \PR{D64} (2001) 065012 
  [{\tt hep-th/0012158}]. 


 \bibitem{AK} H.~Abe and K.~Choi, \JH{0412} (2004) 069 
  [{\tt hep-th/0412174}]. 

 \bibitem{BKV} E.~Bergshoeff, R.~Kallosh and A.~Van Proeyen, 
  \JH{0010} (2000) 033 [{\tt hep-th/0007044}]. 

 \bibitem{MO} N.~Maru and N.~Okada, \PR{D70} (2004) 025002 
  [{\tt hep-th/0312148}]. 

 \bibitem{GW} W.D.~Goldberger and M.B.~Wise, \PRL{83} (1999) 4922 
  [{\tt hep-ph/9907447}]. 

 \bibitem{AS2} H.~Abe and Y.~Sakamura, \PR{D71} (2005) 105010 
  [{\tt hep-th/0501183}]. 

 \bibitem{AS3} H.~Abe and Y.~Sakamura, {\tt hep-th/0508176}. 

 \bibitem{BNZ} J.~Bagger, D.~Nemeschansky and R.~Zhang, 
  \JH{0108} (2001) 057 [{\tt hep-th/0012163}]. 

 \bibitem{SS} J.~Scherk and J.H.~Schwarz, \PL{B82} (1979) 60. 

 \bibitem{AS4} H.~Abe and Y.~Sakamura, \JH{0602} (2006) 014 
 [{\tt hep-th/0512326}]. 
\end{thebibliography}
\end{document}